\documentclass[10pt,english,showpacs,twocolumn,nofootinbib]{revtex4}
\usepackage[latin9]{inputenc}
\usepackage{graphicx}
\usepackage{amsmath}
\usepackage{amssymb}
\usepackage{amsfonts}
\usepackage{lscape}
\usepackage{url}
\usepackage{epsfig}
\usepackage{color}
\usepackage{bm}
\usepackage{babel}
\usepackage{xcolor}
\usepackage{tabularx}



\newcommand{\nn}{\nonumber}
\newcommand{\newc}{\newcommand}
\newc{\be}{\begin{equation}}
\newc{\ee}{\end{equation}}
\newc{\ba}{\begin{eqnarray}}
\newc{\ea}{\end{eqnarray}}
\newc{\bea}{\begin{eqnarray*}}
\newc{\eea}{\end{eqnarray*}}
\def \lcdm {$\Lambda$CDM }

\def \om {\Omega_{{\rm m}_0}}
\def \oma {\Omega_{\rm m}}

\def \Geff {G_{\rm eff}}

\def \fs8 {$f\sigma_8$ }

\newc{\GB}{\mathcal{G}}
\newc{\cs}{c_{s}^2}

\begin{document}
\title{Accuracy of the growth index in the presence of dark energy perturbations}

\author{Savvas Nesseris$^{1}$}
\email{savvas.nesseris@unige.ch}

\author{Domenico Sapone$^{2}$}
\email{dsapone@ing.uchile.cl}

\affiliation{$^{1}$ D\'epartement de Physique Th\'eorique \& Center for Astroparticle Physics, Universit\'e de Gen\`eve, Quai E.\ Ansermet 24, CH-1211 Gen\`eve 4, Switzerland, \\
$^{2}$Cosmology and Theoretical Astrophysics group, Departamento de F\'isica, FCFM, Universidad de Chile,\\
Blanco Encalada 2008, Santiago, Chile}


\begin{abstract}
We present the analytical solutions for the evolution of matter density perturbations, for a model with a constant dark energy equation of state $w$ but when the effects of  the dark energy perturbations are properly taken into account. We consider two cases, the first when the sound speed of the perturbations is zero $c_s^2=0$ and the general case $0<c_s^2 \leq 1$. In the first case our solution is exact, while in the second case we found an approximate solution which works to better than $0.3\%$ accuracy for $k>10 H_0$ or equivalently $k/h>0.0033 \textrm{Mpc}^{-1}$. We also estimate the corrections to the growth index $\gamma(z)$, commonly used to parametrize the growth-rate. We find that these corrections due to the DE perturbations affect the growth index $\gamma$ at the  $3\%$ level. We also compare our new expressions for the growth index with other expressions already present in the literature and we find that the latter are less accurate than the ones we propose here. Therefore, our analytical calculations are necessary as the theoretical predictions for the fundamental parameters to be constrained by the upcoming surveys need to be as accurate as possible, especially since we are entering in the precise cosmology era where parameters will be measured to the percent level.
\end{abstract}
\maketitle

\section{Introduction}
Under the assumptions that the Universe at large scales is homogeneous and isotropic, that it can be described by General Relativity or some other modified gravity theory, such as $f(R)$ or any other metric theory whose effects at the perturbations level can be taken into account by the effective Newtonian constant $\Geff(a)$, see Refs. \cite{Amendola:2007rr, Tsujikawa:2007gd, Nesseris:2008mq, Nesseris:2009jf}, and finally under the subhorizon approximation ($k\gg aH$), then it can be shown that the growth of matter is governed by the second order differential equation:
\be
\delta''(a)+\left(\frac3a + \frac{H'(a)}{H(a)}\right)\delta'(a)-\frac32 \frac{\Omega_{m_0} \Geff(a)/G_N \delta(a)}{a^5 H(a)^2/H_0^2}=0,
\label{ode}
\ee
where $\delta=\frac{\delta \rho_m}{\rho_m}$ is the matter density contrast that describes the growth of matter (known as the growth factor), $H(a)$ is the Hubble parameter, $\Omega_{m_0}$ is the matter density today and $H_0$ is the Hubble constant.

Making the assumption that the dark energy component can be described by a constant equation of state $w$ and negligible dark energy perturbations, i.e. $\Geff(a)/G_N=1$, then Eq.~(\ref{ode}) can be easily solved analytically. The differential equation (\ref{ode}) has in general two solutions that correspond to two different physical modes, a decaying and a growing one, that in a matter dominated Universe in GR behave as $\delta=a^{-3/2}$ and as $\delta=a$ respectively. Since we are only interested in the latter, we demand that at early times $a_{in}\ll1$, usually during matter domination, the initial conditions have to be chosen as $\delta(a_{in})\simeq a_{in}$ and $\delta'(a_{in})\simeq1$. When $\Geff(a)/G_N=1$ we get GR as a subcase, while in general for modified gravity theories, the term $\Geff$ can be scale and time dependent.

For a flat GR model with a constant dark energy equation of state $w$, the exact solution of Eq.~(\ref{ode}) for the growing mode, neglecting the dark energy perturbations, is given by \cite{Belloso:2011ms,Silveira:1994yq, Percival:2005vm}
\ba
\delta(a)&=& a~{}_2F_1\left[- \frac{1}{3 w},\frac{1}{2} -
\frac{1}{2 w};1 - \frac{5}{6 w};a^{-3 w}(1 - \Omega_{m_0}^{-1})\right]
\label{Da1} \nn \\ \textrm{for}&&H(a)^2/H_0^2= \Omega_{m_0} a^{-3}+(1-\Omega_{m_0})a^{-3(1+w)},\ea
where ${}_2F_1(a,b;c;z)$ is a hypergeometric function, see Ref.~\cite{handbook} for more details. In more general cases, for instance admitting that the dark energy equation of state parameter is a function of time, it is impossible to find a closed form analytical solution for Eq.~(\ref{ode}).

However, when we take into account the effect of the dark energy perturbations, even though the effect is small, the analytical solution (\ref{Da1}) is no longer valid. The reason for this is that, as shown in Ref.~\cite{Sapone:2009mb}, the dark energy perturbations can be included effectively as a $\Geff$ in Eq.~(\ref{ode}), given by:
\be
\Geff(a,k)/G_N=1+\frac{1-\Omega_{m_0}}{\Omega_{m_0}}(1+w) \frac{a^{-3w}}{1-3w+\frac{2k^2 c_s^2 a}{3H_0^2\Omega_{m_0}}}\,.\label{geffde}
\ee
Depending on the values of the equation of state parameter $w$ and sound speed $\cs$ the density contrast of the dark energy can be of the order of $\Delta_{de}\sim 0.06 \Delta_m$  and of the order of $\Delta_{de}\sim 0.01 \Delta_m$, for $w=-0.8$ and $w=-0.95$ respectively (of course if $w=-1$ we are dealing with a cosmological constant and hence, by definition it has no perturbations).

As mentioned before, the dark energy fluid considered in this paper should be thought as an effective dark energy component, so the values of the equation of state $w$ and sound speed $\cs$ are effective parameters, hence not always they have a physical meaning but rather they are a representation of the particular model taken into consideration. In practice, considering for instance a value of $\cs =0$ does not necessarily means that we are dealing with a dark energy fluid that behaves like dark matter but it might be due to a particular modified gravity model expressed in terms of fluid parameters which has a value of $\cs$ equal to zero (or very small), see \cite{KunzCA}. It is also worth mentioning that the Eq.~(\ref{geffde}) has been evaluated under the condition of constant in time sound speed, however the sound speed can depend on the scale.

The paper is organized as followed: in Section \ref{sec:solutions} we find exact analytical expressions for the evolution of the matter density contrast sourced by the dark energy perturbations and in Section \ref{sec:comparison} we compare them to numerical solutions and test their accuracy; in Section \ref{sec:growth-index} we find the new growth index $\gamma$ and we compare it with other parameterizations, while in Section \ref{sec:conclusions} we present our conclusions.

\section{The solution for the matter density contrast \label{sec:solutions}}

In what follows we will find the analytic solution to the matter density contrast when dark energy perturbation are properly taken into account. To solve the second order differential equation for $\delta_{m}$ we will consider two different limits: first when the sound speed is effectively zero ($\cs =0$), and second when the sound speed is $0<\cs\leq 1$.

We start by rewriting Eq.~(\ref{ode}) in terms of $G(a)$ defined as $\delta(a) \equiv a G(a)$, the dimensionless Hubble parameter $E(a)\equiv H(z)/H_0$ and the parameter $Q(a,k) \equiv \Geff(a,k)/G_N$, so Eq.~(\ref{ode}) becomes
\ba
&&a^2G''(a)+\left[5+a\frac{E'(a)}{E(a)}\right]aG'(a)\nn\\
&&+\left[3+a\frac{E'(a)}{E(a)}-\frac{3}{2}\frac{\Omega_{m_0}}{a^3E(a)^2}Q(a)\right]G(a)=0\,.
\label{eq:ode-G}
\ea
We need to express the function $Q(a,k)$ which accounts for the dark energy perturbations\footnote{we would like to remind the reader the $G_{eff}$ and $Q$ are the same, hence we use them interchangeably}. In this paper we use the expressions found in \cite{saponeMB}. In the latter the authors solved analytically the full system of differential equations for dark matter and dark energy. In order to find analytic solutions to matter and dark energy density contrasts, some assumptions had to be made, in particular that the equation of state parameter $w$ and the sound speed of the dark energy component have to be constant in time or at least slowly varying (however the sound speed can depend on the scale $k$). Once the analytic solutions for matter and dark energy were found then it was possible to express the quantity $Q(a,k)$ which is defined as the ratio
\be
Q(a, k) = 1+\frac{\rho_{DE}\Delta_{DE}}{\rho_{m}\Delta_{m}}\,.
\ee
which is a phenomenological function that takes into account the relative growth of the dark energy perturbations. This function enters directly into the gravitational potential $k^2\phi = -4\pi G a^2Q(a,k)\rho_m\Delta_m$, that is the reason why $Q(a,k)$ enters only in third term into Eq.~(\ref{eq:ode-G}). 

Another interesting issue is the initial conditions for matter and dark energy. In our case dark energy perturbations are sourced by the dark matter perturbations via the gravitational potential; because we are in linear order perturbation theory then all $k$-modes evolve independently and as a consequence each $k$-mode depends linearly on a normalization factor which is constant in time however it depends on $k$, say $\delta_0(k)$. The value of this constant is given by inflation at very early times. In this scenario, setting the initial conditions means to set the constant $\delta_0$. However, as stated, the factor $\delta_0$ enters as multiplicative factor to both dark matter and dark energy, hence the quantity $Q(a,k)$, which is the main concern of this paper, has no direct dependence on $\delta_0$. A detailed analysis on the initial conditions for the dark energy perturbations and the decaying modes of the solutions can be found in \cite{saponeMB}. The general solution, see Ref.~\cite{saponeMB}, is:
\ba
Q(a,k) &=& 1+ \nn \frac{1-\Omega_{m_0}}{\Omega_{m_0}}(1+w) \frac{a^{-3w}}{1-3w+\frac{2k^2 c_s^2 a}{3H_0^2\Omega_{m_0}}}
\label{eq:q-below1}
\ea
where $\cs$ is the sound speed of dark energy. 

\subsection{The case $\cs=0$}

When the dark energy sound speed is equal to zero, then the $Q(a,k)$ parameter which gives us the amount of the dark energy perturbations can be written as
\be
Q= 1+Q_0a^{-3w}
\ee
where
\be
Q_0 = \frac{1-\om}{\om}\frac{1+w}{1-3w}\,.
\ee
Before inserting the above equation into Eq.~(\ref{eq:ode-G}), we can make the following change of variables to simplify the problem even more.
Let us consider the new variable
\be
u = \frac{1-\om}{\om}a^{-3w}\,\rightarrow\,a\frac{{\rm d}}{{\rm d}a}=-3wu\frac{{\rm d}}{{\rm d}u}
\ee
then the Eq.~(\ref{eq:ode-G}) becomes
\ba
&&u^2\ddot{G}+\left[1-\frac{5}{6w}+\frac{1}{2}\frac{u}{1+u}\right]\frac{u}{3w}\dot{G} +\nn\\
&&+\frac{1}{6w^2}\frac{1-w-q_0}{1+u}u G=0\,
\label{eq:ode-G-u}
\ea
where $q_0 = (1+w)/(1-3w)$. Eq.~(\ref{eq:ode-G-u}) can be solved analytically and the solution is
\ba
G(a) &=& \,_2F_1\left[\frac14-\frac{5}{12w}+B,\frac14-\frac{5}{12w}-B,1-\frac{5}{6w}; \right. \nn\\
&-&\left.\frac{1-\om}{\om}a^{-3w}\right] \label{eq:solcs2-0}
\ea
where
\be
B=\frac{1}{12w}\sqrt{\left(1-3w\right)^2+24\frac{1+w}{1-3w}}\,.
\label{eq:B1-solcs2-0}
\ee

\subsection{The case $\cs\neq 0$}

If the dark energy sound speed is not zero, then the dark energy will have a sound horizon below which perturbations cannot grow. The modification to the Newtonian potential will be, see \cite{saponeMB}:
\ba
Q(a)-1 &=& \nn \frac{1-\om}{\om}(1+w) \frac{a^{-3w}}{1-3w+\frac{2k^2 c_s^2 a}{3H_0^2\om}}\\
&\simeq& (1-\om)\frac32 \left(1+w\right)\frac{H_0^2}{c_s^2 k^2}a^{-1-3w}\,,
\label{eq:q-below}
\ea
where we have used the fact that we work in the subhorizon approximation ($k\gg aH$).
As it can be seen, the $Q(a)$ term is suppressed by a term $a^{-1}$ the lack of growth of the dark energy perturbations when they enter the sound horizon.
However, inserting Eq.~(\ref{eq:q-below}) into Eq.~(\ref{eq:ode-G}) and making the change of variable $a\rightarrow u$, we cannot find an analytic solution to the matter density contrast because of the extra dependence $a^{-1}$ in the expression of $Q(a,k)$. In order to solve Eq.~(\ref{eq:ode-G}) we make the approximation:
\be
Q(a) =1+ Q_0a^{-1-3w}\simeq 1+Q_0 a^{-3w}\,,
\ee
where $Q_0 = (1-\om)\frac32 \left(1+w\right)\frac{H_0^2}{c_s^2 k^2}$; then the differential equation that we need to solve will look exactly as Eq.~(\ref{eq:q-below}) where now we have a new $q_0$ for modes below the sound horizon: $q_0=\frac32 \left(1+w\right)\frac{H_0^2\om}{c_s^2 k^2}$. In this case we are overestimating the amount of dark energy perturbations below the sound horizon as we are taking out the term $a^{-1}$ which lowers the clustering of the perturbations.

Then the solution to the Eq.~(\ref{eq:q-below}) is
\ba
G(a) &=& \,_2F_1\left[\frac14-\frac{5}{12w}+B,\frac14-\frac{5}{12w}-B,1-\frac{5}{6w}; \right. \nn\\
&-&\left.\frac{1-\om}{\om}a^{-3w}\right] \label{eq:solcs2-1}
\ea
where
\be
B= \frac{1}{12w}\sqrt{\left(1-3w\right)^2+36\om\left(1+w\right)\frac{H_0^2}{k^2 c_s^2}}\,.
\label{eq:B1-solcs2-1}
\ee

\subsection{The growth-rate}

In the previous section we have found the solution to the matter density contrast in terms of the hypergeometric function:
\bea
\delta(a) &=& a \,_2F_1\left[\frac14-\frac{5}{12w}+B,\frac14-\frac{5}{12w}-B,1-\frac{5}{6w}; \right. \nn\\
&&\left.1-\frac{1}{\oma(a)}\right]
\label{eq:solcs2-10}
\eea
where
\be
B = \frac{1}{12\,w}\sqrt{(1-3w)^2+24\delta B}\,
\ee
and the parameter $\delta B$ can accommodate  all different cases if defined as
\ba
\delta B &=& 0~~~~~~~~~~~~~~~~~~~~~~~~\textrm{(No DE perts)},\\
\delta B &=& \frac{(1+w)}{1-3 w}~~~~~~~~~~~~~~~(\textrm{for}~c_s^2=0),\\
\delta B &=& \frac{36}{24} \om (1+w) \frac{H_0^2}{k^2 c_s^2}~~(\textrm{for}~c_s^2>0).
\ea
Then, the growth rate $f(a)$ is
\bea
f(a) &=& a\frac{\delta'(a)}{\delta(a)} =1 +\nonumber \\
&+&a\frac{\oma'(a)}{\oma^2(a)}\frac{\alpha\beta}{\gamma}\frac{\,_2F_1\left[\alpha+1, \beta+1, \gamma+1;1-\frac{1}{\oma(a)}\right]}{\,_2F_1\left[\alpha, \beta, \gamma;1-\frac{1}{\oma(a)}\right]}
\eea
where the coefficients are
\bea
\alpha &\equiv & \frac14-\frac{5}{12w}+B\,,\\
\beta  &\equiv & \frac14-\frac{5}{12w}-B\,,\\
\gamma &\equiv & \frac12+\alpha+\beta\,.
\eea
It is interesting to notice that the hypergeometric function can be simplified making use of the relation between hypergeometric functions and Legendre polynomial $P_\nu^\mu(x)$. Using Eqs.~(15.4.12) and (15.4.21) and Eqs.~(8.5.3) and (8.5.5) of Ref.~\cite{handbook} we find
\be
f(a) = 1+3w\alpha\left(1-\sqrt{\oma(a)}\frac{P_{2B+\frac12}^{5/6w}\left[1/\sqrt{\oma(a)}\right]}{P_{2B-\frac12}^{5/6w}\left[1/\sqrt{\oma(a)}\right]}\right).
\label{eq:growthrate-legendre}
\ee
If dark energy perturbations are switched to zero we have
\be
f(a) = \sqrt{\oma(a)}\frac{P_{1/6w}^{5/6w}\left[1/\sqrt{\oma(a)}\right]}{P_{-1/6w}^{5/6w}\left[1/\sqrt{\oma(a)}\right]}\,,
\ee
in agreement with the expression found in Ref.~\cite{Belloso:2011ms}.

\subsection{Joint solution}

Alternatively, we can also find a joint solution to the equations in order to have one single solution. The difference in the two solutions comes in the coefficients of the hypergeometric functions which then account for the dark energy perturbations, i.e. the coefficients $B$ in Eqs.~(\ref{eq:B1-solcs2-0}) and (\ref{eq:B1-solcs2-1}). Hence we can think to join these two coefficients directly. We found that
\ba
B_{c_s^2>0} &=& \frac{1}{12w} \sqrt{(1-3w)^2+36\om(1+w)\frac{H_0^2}{k^2\cs}} \nonumber \\
&=&\frac{1}{12w}\sqrt{(1-3w)^2+\delta B_1} \\
B_{c_s^2=0}&=& \frac{1}{12w}\sqrt{(1-3w)^2+24\frac{1+w}{1-3w}}\nonumber \\
&=& \frac{1}{12w}\sqrt{(1-3w)^2+\delta B_2}
\ea
for scales above and below the sound horizon, respectively. Joining only the parts that account for dark energy perturbations we find
\ba
&&B_{joint} = \frac{1}{12w}\sqrt{(1-3w)^2+\frac{\delta B_1\delta B_2}{\delta B_1+\delta B_2}} = \nonumber\\
&&= \frac{1}{12w} \sqrt{(1-3w)^2+24\frac{1+w}{1-3w+\frac{2}{3}\frac{k^2}{H_0^2\om}\cs}}
\label{eq:total-b}
\ea
We can also add an $a$ dependence to the coefficient. The reason is that the second term in the square root is exactly $Q$ used, hence the dark energy perturbations, and we have
\ba
&&B_{joint} = \frac{1}{12w}\sqrt{(1-3w)^2+\frac{\delta B_1\delta B_2}{\delta B_1+\delta B_2}} = \nonumber\\
&& = \frac{1}{12w} \sqrt{(1-3w)^2+24\frac{1+w}{1-3w+\frac{2}{3}\frac{a\,k^2}{H_0^2\om}\cs}}\,.
\label{eq:total-b-a}
\ea
However, in this paper we consider Eq.~(\ref{eq:total-b}) for two main reasons: the presence of the scale factor does not change the solution for $f\sigma_8$ (the difference between the two is of the order of $10^{-8}$, and second it is not really clear how to deal, in the hypergeometric function, with coefficients that depend on the variable.

The expression found will be very useful for instance to speed up the code for forecasts, analytic estimates of $\gamma$ or fitting the data. The growth index $\gamma$ will be shown in the next section to be the same, except that we have to use now the new $\delta B = 24\frac{1+w}{1-3w+\frac{2}{3}\frac{k^2}{H_0^2\om}\cs}$.

\section{The growth index \label{sec:growth-index}}
Here we present the corrections to the growth index $\gamma(a)$ due to the dark energy perturbations. In Ref.~\cite{Wang:1998gt} it was shown that the growth rate $f(a)\equiv \frac{d ln \delta}{dlna}$, in the case of no DE perturbations, can be approximated as
\ba
f(a)&=&\oma(a)^{\gamma(a)} \label{fg}\\
\om(a)&\equiv&\frac{\om~a^{-3}}{H(a)^2/H_0^2} \\
\gamma(a) &=&\frac{\ln f(a)}{\ln \oma(a)}\simeq \frac{3 (1-w)}{5-6 w}+\cdots \label{eq:wst1}
\ea
When we want to include the DE perturbations, one way to do it is via the semianalytic approach of Ref. \cite{Linder:2007hg}, where it was shown that the growth index depends on $\Geff$ as
\ba
\gamma&=&\frac{3(1-w-A(\Geff))}{5-6w} \label{linder}, \\
A(\Geff)&=&\frac{\Geff-1}{1-\oma(a)}.
\ea
For other approaches that explore the effects of the dark energy perturbations or modified gravity in general on the growth index, see Ref.~\cite{Basilakos:2014yda,Mehrabi:2015hva,Dossett:2013npa,Piattella:2014lba,Burrage:2015lla}.
However, here we will follow a more straight-forward approach by using the analytic expressions found in the previous sections. 
In what follows we will use the shorthand of $\Omega$ to mean $\om(a)$ in order to simplify the notation. Then using the ansatz $f(\Omega)=\Omega^{\gamma(\Omega)}$  we find that the growth index up to first order can be written as
\ba
&&\gamma = \frac{\ln(f(\Omega))}{\ln(\Omega)}  \nn \\
&&=\frac{3 (\delta B+w-1)}{6 w-5}-\nn \\
&&\frac{3 (\Omega -1) ((\delta B+w-1) (9 \delta B (4 w-3)-3 w+2))}{2 \left((5-6 w)^2 (12 w-5)\right)} + \cdots \nn \\
&&\,
\label{eq:new-gamma} \ea
The reason we perform the series expansion of $\gamma$ in terms of $\Omega$ is that it allows us to extract the zero-th order part of $\gamma$, which in the $\Lambda$CDM model is 6/11, but also the first order correction. As a result we can compare with the other expressions commonly found in the literature, eg Eq.~(\ref{eq:wst1}), and compare their accuracy. 

It is instructive to split the contributions to the growth index to two parts:
\be
\gamma=\gamma_m+\gamma_{DE}, \label{eq:gammaeq}
\ee
where
\ba
&&\hspace{-0.4cm}\gamma_m = \frac{3 (w-1)}{6 w-5}+ \frac{3 (3 w-2) (w-1) (\Omega -1)}{2 (5-6 w)^2 (12 w-5)} +\cdots,\\
&&\hspace{-0.4cm}\gamma_{DE}= \frac{3 \delta B}{6 w-5}+(\Omega-1)\left( -\frac{3 \delta B (6 w (6 w-11)+29)}{2 \left((5-6 w)^2 (12 w-5)\right)}-\right. \nn \\
&& \left. \frac{27 \delta B^2 (4 w-3)}{2 \left((5-6 w)^2 (12 w-5)\right)}\right)+\cdots,
\ea
where the first term $\gamma_m$ corresponds to the well known result from Ref.~\cite{Wang:1998gt} and the second term $\gamma_{DE}$ corresponds to the extra contribution of the DE perturbations.

Alternatively, we can follow the procedure of Ref.~\cite{Wang:1998gt} to evaluate the growth rate. We can do this by changing variables from $a$ to $\Omega$ and replacing the growth factor $\delta(a)$ with the growth rate $f(\Omega)=a\frac{\delta'(a)}{\delta(a)}$.  Doing that we arrive at a differential equation for $f(\Omega)$ in terms of $\Omega$:
\ba
&& 3 w (1-\Omega ) \Omega  f'(\Omega )+f(\Omega ) \left(\frac{1}{2}-\frac{3}{2} w (1-\Omega )\right)+\nn \\ &&f(\Omega )^2-\frac{3}{2} \Omega  \Geff(\Omega )=0.
\ea
Using the ansatz $f(\Omega)=\Omega^{\gamma(\Omega)}$  we arrive to the same result as in Eq.~(\ref{eq:gammaeq}).

\subsection{The case $\cs = 0$}
In the case when the sound speed is zero, we can use Eq.~(\ref{eq:gammaeq}) with $\delta B = \frac{(w+1)}{1-3 w}$ to find
\ba
\gamma(\Omega)&=& \frac{3 w (3 w-5)}{(3 w-1) (6 w-5)}+\nn \\&&\frac{15 w (3 w-5) \left(9 w^2-5\right)}{2 (12 w-5) \left(18 w^2-21 w+5\right)^2}(\Omega-1)\cdots \nonumber \\
~~~~\label{gammaDE1}
\ea

Also, we separate the effects of the matter and dark energy perturbations on $\gamma$ as
\be
\gamma=\gamma_m+\gamma_{DE},
\ee
where
\ba
\gamma_m &=& \frac{3 (w-1)}{6 w-5}+ \frac{3 (3 w-2) (w-1) (\Omega -1)}{2 (5-6 w)^2 (12 w-5)}, \label{eq:gammam1}\\
\gamma_{DE}&=& -\frac{3 (w+1)}{18 w^2-21 w+5}+\nn \\ &&(\Omega -1)\frac{3 (w+1) (9 w (3 w (2 w-5)+8)-1) }{(12 w-5) \left(18 w^2-21 w+5\right)^2}.\nn  \\
&& \label{eq:gammade1}
\ea
We can also compute the ratio of the two quantities $\gamma_{DE}/\gamma_m$ in order to estimate the effect of neglecting the DE perturbations on the measurement of $\gamma$ in up-coming surveys. We find that the ratio can be written as
\ba
\frac{\gamma_{DE}}{\gamma_m}&=&-\frac{w+1}{3 w^2-4 w+1}+\nn\\&&\frac{9 w (w+1) (3 w-5) (4 w-3) (\Omega -1)}{2 (1-3 w)^2 (w-1) (6 w-5) (12 w-5)}+\cdots . \nn\\
\label{ratiogammas}
\ea

\begin{figure}[t!]
\centering
\vspace{0cm}\rotatebox{0}{\vspace{0cm}\hspace{0cm}\resizebox{0.48\textwidth}{!}{\includegraphics{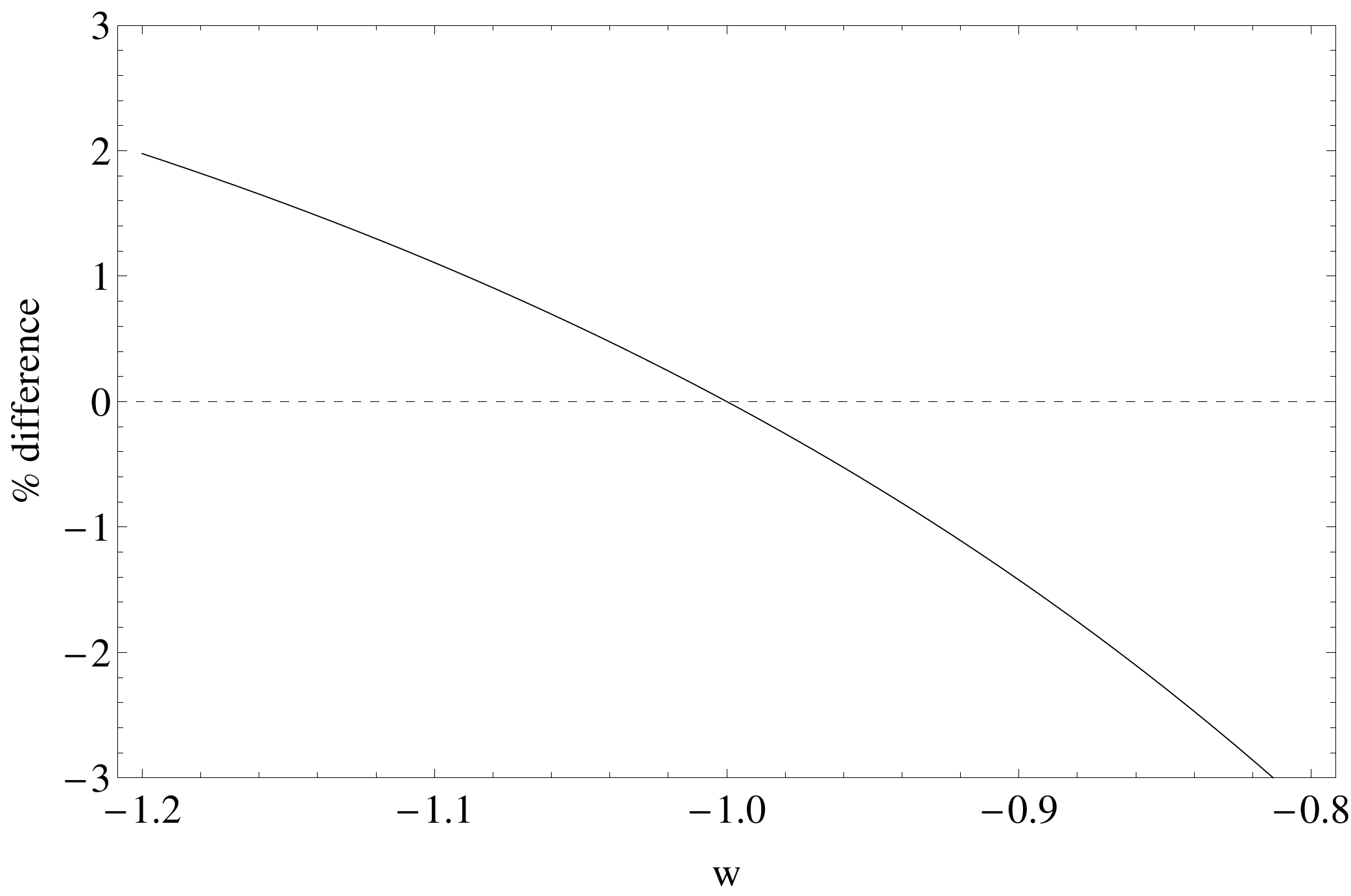}}}
\caption{The percent difference of the effect of the DE perturbations of $\gamma$ as part of the total and as a function of the equation of state $w$ for $c_s^2=0$. In practice we plot $100*\frac{\gamma_{DE}}{\gamma_m}$, with the latter given by the zero order part of Eq.~(\ref{ratiogammas}).\label{fig:ratio}}
\end{figure}

\begin{figure}[t!]
\centering
\vspace{0cm}\rotatebox{0}{\vspace{0cm}\hspace{0cm}\resizebox{0.48\textwidth}{!}{\includegraphics{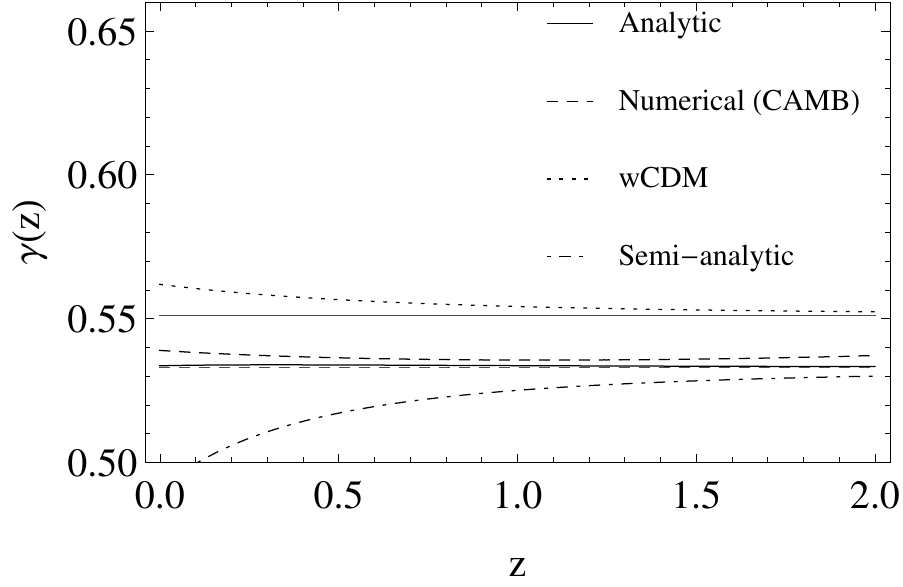}}}
\caption{A comparison of all the different expressions for $\gamma$ for $(c_s^2,w,\om)=(0,-0.8,0.3)$. The solid black line corresponds to the expression for the analytic solution of Eq.~(\ref{gammaDE1}), the dashed black line to the numerical solution from CAMB, the dotted line to the wCDM model with no DE perturbations, the dot-dashed black line to the semianalytic expression of Eq.~(\ref{linder}), while the red and dashed-red lines correspond to the zero-order terms for expansions for $\gamma$ without and with DE perturbations respectively.  \label{fig:gammas}}
\end{figure}

In Fig. \ref{fig:ratio} we show the percent difference of the effect of the DE perturbations as part of the total and as a function of the equation of state $w$. In practice we plot $100*\frac{\gamma_{DE}}{\gamma_m}$, with the latter given by Eq.~(\ref{ratiogammas}). For $w=-1$ Eq.~(\ref{gammaDE1}) gives the expected result $\gamma=6/11$, but for $w\neq-1$ there are substantial differences, eg for $w=-0.8$ we have $\gamma=0.533$ which is a $3.4\%$ difference from the result $\gamma=0.551$ found when neglecting DE perturbations. These differences can be important as cosmology has moved into a high precision era and percent accuracies will be sought after in the cosmological parameters by the upcoming surveys like DES, LSST and Euclid, where the relative errors on the growth index $\gamma$ are of the order of  $0.5-3\%$ depending if combined probes are used, see \cite{Belloso:2011ms}, \cite{Abell:2009aa}, \cite{AmendolaYS}, \cite{HutererXKY}.

In Fig. \ref{fig:gammas} we show a comparison of all the different expressions for $\gamma$ for $(c_s^2,w,\om)=(0,-0.8,0.3)$. The solid black line corresponds to the expression for the analytic solution of Eq.~(\ref{gammaDE1}), the dashed black line to the numerical solution from CAMB, the dotted line to the wCDM model with no DE perturbations, the dot-dashed black line to the semianalytic expression of Eq.~(\ref{linder}), while the red and dashed-red lines correspond to the zero-order terms for expansions for $\gamma$ without and with DE perturbations respectively. 

Clearly, our analytic solution in this case is in excellent agreement with the numerical one and by far superior to the semianalytic one. Specifically, the expression of Eq.~(\ref{linder}), which is the dot-dashed curve in Fig. \ref{fig:gammas}, differs by more than $\sim 6\%$ at $z\sim0$. The reason for this is that it is a semianalytic approximation derived at high $z$ and as a result at low redshifts it is significantly different from both the exact and numerical solutions. Therefore, we find that it is not suitable for use in Fisher Matrix analysis or data fitting when the DE perturbations are taken into account.

Finally, we should mention that we have checked that using the growth index $\gamma$ of Eq.~(\ref{eq:new-gamma}) to calculate the growth-rate $f$ and $f\sigma_8$ by using and integrating Eq.~(\ref{fg}) respectively, is in agreement to better than $0.1\%$ compared to using the full analytical solution of Eq.~(\ref{eq:solcs2-0}).

\subsection{The case $\cs\neq 0$}
When the sound speed $c_s^2$ is different from zero, we can again use Eq.~(\ref{eq:gammaeq}) with $\delta B = \frac{36}{24} \kappa (1+w)$, where we have set $\kappa \equiv \frac{\Omega_{m,0} H_0^2}{k^2 c_s^2}$, to find
\ba
\gamma(\Omega)&=& \frac{3 (3 \kappa +(3 \kappa +2) w-2)}{2 (6 w-5)}- \frac{3(\Omega -1)}{8 (5-6 w)^2 (12 w-5)}\cdot\nn \\&&\left(27 \kappa  \left(4 w^2+w-3\right)-6 w+4\right) \cdot \nn \\
&&(3 \kappa +(3 \kappa +2) w-2)+\cdots. \label{gammaDEcs21}
\ea
As mentioned earlier in this section and in order to simplify the notation, by writing  $\Omega$ we actually mean $\om(a)$ and $\Omega_{m,0}\equiv \om(a=1)$.

Also, we can separate the effects of the matter and dark energy perturbations on $\gamma$ as
\be
\gamma=\gamma_m+\gamma_{DE},
\ee
where
\ba
\gamma_m &=& \frac{3 (w-1)}{6 w-5}+ \frac{3 (3 w-2) (w-1) (\Omega -1)}{2 (5-6 w)^2 (12 w-5)}, \label{eq:gammamcs21}\\
\gamma_{DE}&=& \frac{9 \kappa  (w+1)}{2 (6 w-5)}-\frac{9(\Omega -1)\kappa  (w+1)}{8 (5-6 w)^2 (12 w-5)}\cdot\nn \\ && \left(27 \kappa  \left(4 w^2+w-3\right)+12 w (6 w-11)+58\right).\nn  \\
&& \label{eq:gammadecs21}
\ea
We can also compute the ratio of the two quantities $\gamma_{DE}/\gamma_m$ in order to estimate the effect of neglecting the DE perturbations on the measurement of $\gamma$ in up-coming surveys. We find that the ratio can be written as
\ba
\frac{\gamma_{DE}}{\gamma_m}&=&\frac{3 \kappa  (w+1)}{2 (w-1)}-27(\Omega -1)\kappa  (w+1)\cdot\nn\\&&\frac{(4 w-3) (3 \kappa +(3 \kappa +2) w-2)}{8 (w-1) (6 w-5) (12 w-5)}+\cdots . \nn\\
\label{ratiogammas1}
\ea

\section{Comparison with the numerical solution \label{sec:comparison}}

\begin{figure*}[t!]
\centering
\vspace{0cm}\rotatebox{0}{\vspace{0cm}\hspace{0cm}\resizebox{0.48\textwidth}{!}{\includegraphics{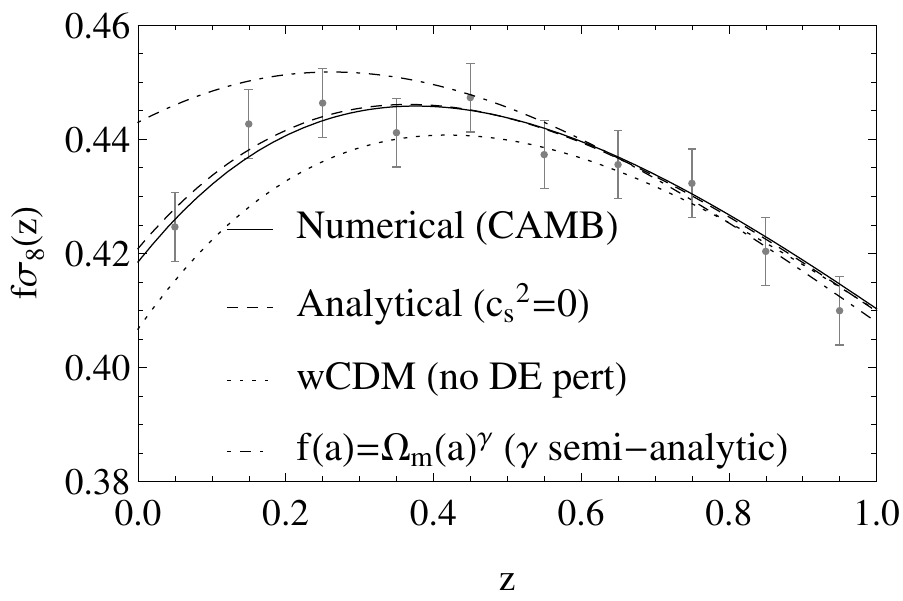}}}
\vspace{0cm}\rotatebox{0}{\vspace{0cm}\hspace{0cm}\resizebox{0.48\textwidth}{!}{\includegraphics{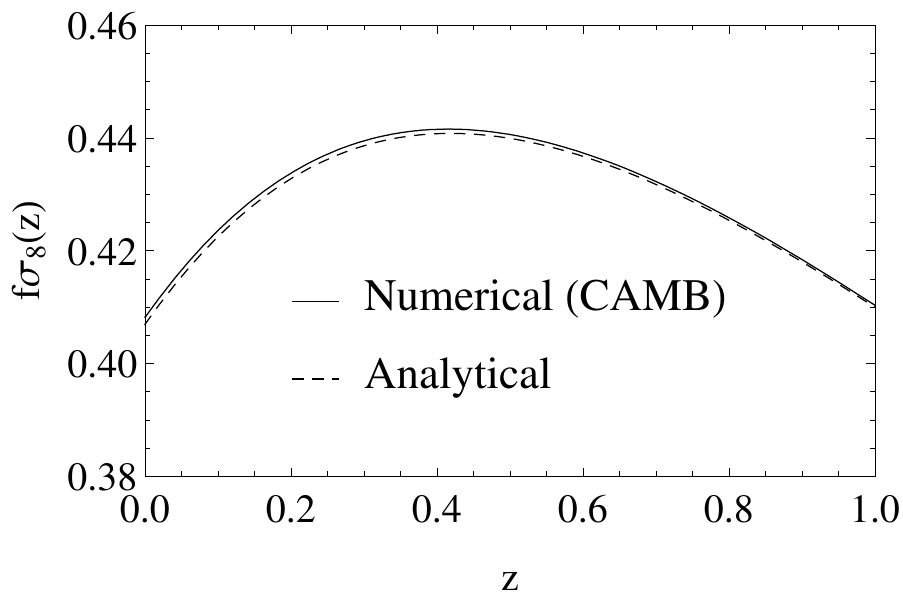}}}
\caption{Left: Comparison of the analytical solution for $f\sigma 8 (z)$ (dashed line) vs the numerical solution of the full system of differential equations for the DE perturbations by CAMB (solid black line), the case with no DE perturbations (dotted line) and the $\oma(a)^\gamma$ parametrization with $\gamma$ given by Eq.~(\ref{linder}) (dot-dashed line) for $c_s^2=0$, $k=200H_0$ and $(w,\om,\sigma_{8,0})=(-0.8,0.3,0.8)$. The gray points correspond to mock $f\sigma8$ data based on the specifications of an LSST-like survey \cite{Abate:2012za}, as it was done in Ref.~\cite{Nesseris:2014qca}. Right: Comparison of the analytical solution (black dashed line) vs the numerical solution (solid black line) of the ODE of Eq.~(\ref{ode}) for $c_s^2=1$ and $k=10H_0$.
\label{fig:comparison1}}
\end{figure*}

\begin{figure}[t!]
\centering
\vspace{0cm}\rotatebox{0}{\vspace{0cm}\hspace{0cm}\resizebox{0.48\textwidth}{!}{\includegraphics{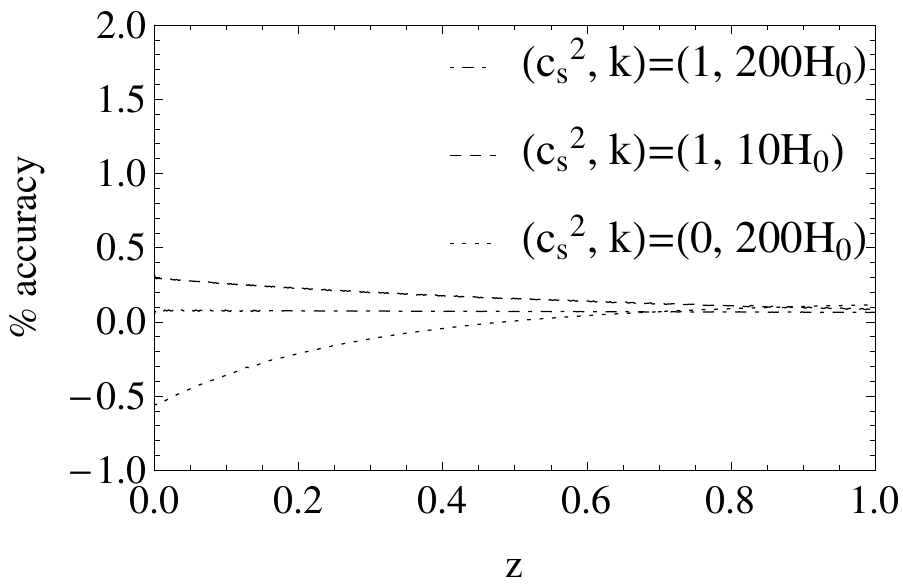}}}
\caption{The percent difference between the analytical solution and the numerical one from CAMB, for various values of the sound speed $c_s^2$ and the scale $k$. \label{fig:comparisonpc}}
\end{figure}

Here we compare our analytical solution  of Eq.~(\ref{eq:solcs2-1}) with the parameter $B$ given by Eqs.~(\ref{eq:B1-solcs2-0}) and Eqs.~(\ref{eq:B1-solcs2-1}), with the numerical one for various choices of parameters and for both cases. Since the data are given in terms of the parameter $f\sigma_8(a)\equiv f(a) \sigma(a)=\frac{\sigma_{8,0}}{\delta(1)} a \delta'(a)$, we will prefer to compare that and the growth rate $f(a)=\frac{d\ln(\delta)}{d\ln a}$.

In Fig.~\ref{fig:comparison1} (left) we show a comparison of the analytical solution for $f\sigma 8 (z)$ (dashed line) vs the numerical solution of the full system of differential equations for the DE perturbations by CAMB (solid black line), the case with no DE perturbations (dotted line) and the $\oma(a)^\gamma$ parametrization with $\gamma$ given by Eq.~(\ref{linder}) (dot-dashed line) for $c_s^2=0$, $k=200H_0$ and $(w,\om,\sigma_{8,0})=(-0.8,0.3,0.8)$. The gray points correspond to mock $f\sigma8$ data based on the specifications of an LSST-like survey \cite{Abate:2012za}, as it was done in Ref.~\cite{Nesseris:2014qca}. As can be seen, the difference between the cases where we include (black solid or dashed lines) or neglect DE perturbations (dotted line) can be significant compared to the small error bars of the expected data from Euclid or LSST and might bias the results.

In Fig.~\ref{fig:comparison1} (right) we show of the analytical solution (black dashed line) vs the numerical solution (solid black line) of the ODE of Eq.~(\ref{ode}) for $c_s^2=1$ and $k=10H_0$. In this case, we also test the effect of the different scale $k$ but also the non-zero sound speed and we see that the difference can be smaller. Our analytical solutions either exact or approximate were found to be in excellent agreement with the numerical ones. In  Fig.~\ref{fig:comparisonpc} we show the percent difference between the analytical solution and the numerical one from CAMB, for various values of the sound speed $c_s^2$ and the scale $k$ and as it can be seen, in all cases we find agreement better than $0.5\%$.

\section{Forecasts}
In this section we test the new expression for the growth index $\gamma$ given by Eq.~(\ref{eq:new-gamma}) in order to constrain the cosmological parameters with the upcoming surveys and we compare the results with the constraints obtained using the growth index given by Eq.~(\ref{linder}). To constrain the cosmological parameters such as the equation of state parameter $w$ and the sound speed $\cs$, we adopt the Fisher matrix technique having in mind a set up similar to LSST experiment \cite{Abell:2009aa}. The LSST is capable of exploring the universe in a range in redshifts of $z \in [0,2]$ covering an area of about $20000\,{\rm deg}^2$. The error on redshift is $\sigma_z = 0.02(1+z)$ which corresponds to the goal of the LSST; the galaxies are distributed with $n(z) \propto z^\alpha\exp\left[-(z/z_1)^\beta\right]$ where $\alpha = 2$, $z_1 = 0.5$ and $\beta = 1$.

In order to avoid non-linearity problems (both in the spectrum and in the bias), we evaluate the Fisher matrix up to a limiting $k_{max}(z)$ at each $z$: we choose values from $0.11h/$Mpc for low-$z$ bins to $0.3h/$Mpc for the highest $z$-bins, see Ref.~\cite{saponeJN,saponeDEP2,Sapone:2013wda} for more details on the Fisher matrix calculations. For the Fisher matrix we consider the following parameters:
$\theta = \left[\omega_m, \omega_b, \tau, n_s, \Omega_{m,0}, w, \cs, P_{shot}\right]$,
where $\omega_m = \om h^2$, $\omega_b = \Omega_{{\rm b}_0}h^2$ and $P_{shot}$ the shot noise. The reference cosmology used in the analysis is the best-fit given by Planck \cite{planck} with the exception of $w = -0.8$ and the sound speed $\cs$ for which we used first a value $\cs =0$ and then $\cs=1$.

For a sound speed of $c_s^2=0$ we find: $\{\sigma_{w},\sigma_{\cs}\} = \{0.0089, 1.67\times 10^{-6}\}$ and $\{\sigma_{w},\sigma_{\cs}\} = \{0.0104, 3.69\times 10^{-6}\}$, for that of Eq.~(\ref{linder}) and our expression respectively. For a sound speed $c_s^2=1$ we find: $\{\sigma_{w},\sigma_{\cs}\} = \{0.0052, 76.63\}$ and $\{\sigma_{w},\sigma_{\cs}\} = \{0.0058, 94.00\}$, for the semianalytic expression and the analytic expression respectively.

Using the expression for the growth index given by Eq.~(\ref{eq:new-gamma}), the errors on the parameters are increased of about $50\%$ for the sound speed and of about $15\%$ for the equation of state parameter, for the $\cs=0$ case. The reason can be found looking at the Fig.~\ref{fig:gammas}. The semianalytic expression for $\gamma(a)$ especially at low redshifts is smaller than the full numerical result: the growth rate $f(a)$ (and also the growth of matter $G(a)$) are proportional to $\Omega_{m}(a)^{\gamma(a)}$ and so do the derivatives with respect to the cosmological parameters. The matter density parameter $\Omega_{m}(a)$ is always lower the unity (it is equal to $1$ only at very high redshifts during matter domination era), consequently if the growth index $\gamma$ is decreased the overall effect on the growth rate is increased and this is reflected in Fig.~\ref{fig:comparison1}. Hence, using the semianalytic expression overestimates (i.e. reduce) the errors.

\section{Conclusions \label{sec:conclusions}}
In Ref. \cite{Sapone:2009mb} it was shown that the effect of the dark energy perturbations on the matter density perturbations is to induce a $\Geff(a,k)$, as shown in Eqs.~(\ref{ode}) and (\ref{geffde}). In this work we have found the analytical solution to Eq.~(\ref{ode}) for constant $w$ when $c_s^2=0 $ and we have explicitly shown that not only is the solution different from the one when the DE perturbations are neglected, see Eq.~(\ref{Da1}), but the difference is actually large enough to affect the results of future surveys like Euclid or LSST, see Fig.~\ref{fig:comparison1} (right). In the case of $0<c_s^2\leq1$ we presented analytic approximations that are in excellent agreement with the numerical solution to Eq.~(\ref{ode}).

Furthermore, we also found analytical expressions for the growth index $\gamma$ used commonly to parameterize the growth as $f(a)=\om(a)^{\gamma}$. This parameter is given by $\gamma\simeq \frac{3 -3 w}{5-6 w}$ for a DE model with constant $w$ in GR when the DE perturbations  are ignored  and is 6/11 for \lcdm ($w=-1$), but when the DE perturbations are properly taken into account it is given by $\gamma\simeq \frac{3 w (3 w-5)}{(3 w-1) (6 w-5)}$ when $c_s^2=0 $, which however again becomes 6/11 for $w=-1$ since \lcdm does not have DE perturbations. 

In the case when the DE sound speed is non-zero we found that the growth index is given by $\gamma\simeq \frac{3 (3 \kappa +(3 \kappa +2) w-2)}{2 (6 w-5)}$, where $\kappa \equiv \frac{\Omega_{m,0} H_0^2}{k^2 c_s^2}$ and has an explicit scale dependence.  Our expression is different but at the same time more accurate than other expressions that have appeared in the literature. We found that it works to better than $0.3\%$ accuracy for $k>10 H_0$ or equivalently $k/h>0.0033 \textrm{Mpc}^{-1}$, thus making it extremely useful for use in forecasts for future surveys. We have also compared our new expressions to that one of Eq.~(\ref{linder}), via a Fisher Matrix approach, and we found that the latter significantly overestimates the errors compared to the more accurate growth index of Eq.~(\ref{eq:new-gamma}) we presented for the first time in this paper.

In conclusion, we have shown that using less accurate expressions for the growth index, such as the one of Eq.~(\ref{linder}), or more importantly neglecting completely the DE perturbations can lead to misleading estimations of the growth index $\gamma$ to the percent level, something which is highly relevant as cosmology today has moved into a high precision era and percent accuracies will be sought after in the next generation surveys.

\section*{Acknowledgments}
The authors would like to thank M.~Kunz and I.~Sawicky for useful discussions. SN acknowledges financial support from the Swiss National Science Foundation.
DS acknowledges financial support from the Fondecyt project number 11140496 and from the ``Anillo'' project ACT1122 founded by the ``Programa de Investigaci\'on asociativa''.


{}


\begin{thebibliography}{}
\bibitem{Amendola:2007rr}
  L.~Amendola, M.~Kunz and D.~Sapone,
  JCAP {\bf 0804} (2008) 013
  [arXiv:0704.2421 [astro-ph]].

\bibitem{Tsujikawa:2007gd}
  S.~Tsujikawa,
  Phys.\ Rev.\  D {\bf 76} (2007) 023514
  [arXiv:0705.1032 [astro-ph]].

\bibitem{Nesseris:2008mq}
  S.~Nesseris,
  Phys.\ Rev.\ D {\bf 79}, 044015 (2009)
  [arXiv:0811.4292 [astro-ph]].

\bibitem{Nesseris:2009jf}
  S.~Nesseris and A.~Mazumdar,
  Phys.\ Rev.\ D {\bf 79}, 104006 (2009)
  [arXiv:0902.1185 [astro-ph.CO]].

\bibitem{Sapone:2009mb}
  D.~Sapone and M.~Kunz,
  Phys.\ Rev.\ D {\bf 80}, 083519 (2009)
  [arXiv:0909.0007 [astro-ph.CO]].

\bibitem{Belloso:2011ms}
  A.~B.~Belloso, J.~Garcia-Bellido and D.~Sapone,
  JCAP {\bf 1110} (2011) 010
  [arXiv:1105.4825 [astro-ph.CO]].

\bibitem{Silveira:1994yq}
  V.~Silveira and I.~Waga,
  Phys.\ Rev.\  D {\bf 50}, 4890 (1994).

\bibitem{Percival:2005vm}
  W.~J.~Percival,
  Astron.\ Astrophys.\  {\bf 443}, 819 (2005)
  [arXiv:astro-ph/0508156].

\bibitem{handbook}
  Abramowitz, Milton; Stegun, Irene A., eds. (1972), \textit{Handbook of Mathematical Functions with Formulas, Graphs, and Mathematical Tables}, New York: Dover Publications, ISBN 978-0-486-61272-0

\bibitem{KunzCA}
M.~Kunz and D.~Sapone,
Phys.\ Rev.\ Lett.\  {\bf 98}, 121301 (2007).
[astro-ph/0612452].

\bibitem{saponeMB} 
D.~Sapone and M.~Kunz, 
Phys.\ Rev.\ D {\bf 80}, 083519 (2009).
[arXiv:0909.0007 [astro-ph.CO]].


\bibitem{Wang:1998gt}
  L.~M.~Wang and P.~J.~Steinhardt,
  Astrophys.\ J.\  {\bf 508}, 483 (1998)
  [astro-ph/9804015].

\bibitem{Nesseris:2014qca}
  S.~Nesseris, D.~Sapone and J.~Garcia-Bellido,
  arXiv:1410.0338 [astro-ph.CO].

\bibitem{Abate:2012za}
  A.~Abate {\it et al.}  [LSST Dark Energy Science Collaboration],
  arXiv:1211.0310 [astro-ph.CO].

\bibitem{Linder:2007hg}
  E.~V.~Linder and R.~N.~Cahn,
  Astropart.\ Phys.\  {\bf 28}, 481 (2007)
  [astro-ph/0701317].

\bibitem{Basilakos:2014yda} 
  S.~Basilakos,
  Mon.\ Not.\ Roy.\ Astron.\ Soc.\  {\bf 449}, 2151 (2015)
  [arXiv:1412.2234 [astro-ph.CO]].

\bibitem{Mehrabi:2015hva} 
  A.~Mehrabi, S.~Basilakos and F.~Pace,
  arXiv:1504.01262 [astro-ph.CO].
  
\bibitem{Dossett:2013npa} 
  J.~Dossett and M.~Ishak,
  Phys.\ Rev.\ D {\bf 88}, no. 10, 103008 (2013)
  [arXiv:1311.0726 [astro-ph.CO]].
  
\bibitem{Piattella:2014lba} 
  O.~F.~Piattella, D.~L.~A.~Martins and L.~Casarini,
  JCAP {\bf 1410}, no. 10, 031 (2014)
  [arXiv:1407.4773 [astro-ph.CO]].
  
\bibitem{Burrage:2015lla} 
  C.~Burrage, D.~Parkinson and D.~Seery,
  arXiv:1502.03710 [astro-ph.CO].

\bibitem{planck}  P.~A.~R.~Ade {\it et al.}  [Planck Collaboration],
Astron.\ Astrophys.\  (2014).
[arXiv:1303.5076 [astro-ph.CO]].

\bibitem{Abell:2009aa}
  P.~A.~Abell {\it et al.}  [LSST Science and LSST Project Collaborations],
  arXiv:0912.0201 [astro-ph.IM].

\bibitem{saponeJN}
D.~Sapone and L.~Amendola,
[arXiv:0709.2792 [astro-ph]].

\bibitem{saponeDEP2} D.~Sapone, M.~Kunz and L.~Amendola,
Phys.\ Rev.\ D {\bf 82}, 103535 (2010).
[arXiv:1007.2188 [astro-ph.CO]].

\bibitem{Sapone:2013wda}
  D.~Sapone, E.~Majerotto, M.~Kunz and B.~Garilli,
  Phys.\ Rev.\ D {\bf 88} (2013) 043503
  [arXiv:1305.1942 [astro-ph.CO]].
  
\bibitem{AmendolaYS}
  L.~Amendola {\it et al.}  [Euclid Theory Working Group Collaboration],
Living Rev.\ Rel.\  {\bf 16}, 6 (2013).
[arXiv:1206.1225 [astro-ph.CO]].

\bibitem{HutererXKY}
  D.~Huterer, D.~Kirkby, R.~Bean, A.~Connolly, K.~Dawson, S.~Dodelson, A.~Evrard and B.~Jain {\it et al.},
Astropart.\ Phys.\  {\bf 63}, 23 (2015).
[arXiv:1309.5385 [astro-ph.CO]].

 
\end{thebibliography}
\end{document}